# B Decays as a Probe of Spontaneous CP-Violation in SUSY Models

Oleg Lebedev
For the CDF Collaboration

*Virginia Tech*
*Blacksburg, Virginia 24061-0435*

*Fermi National Accelerator Laboratory*
*P.O. Box 500, Batavia, Illinois 60510*

July 1999

Presented at *SUSY'99*,
Fermilab, Batavia, Illinois, June 14-19, 1999

Operated by Universities Research Association Inc. under Contract No. DE-AC02-76CH03000 with the United States Department of Energy

# B Decays as a Probe of Spontaneous CP-Violation in SUSY Models

Oleg Lebedev

*Physics Department, Virginia Tech*
*Blacksburg, VA 24061-0435*

**Abstract:** We consider phenomenological implications of susy models with spontaneously broken CP-symmetry. In particular, we analyze CP-asymmetries in B decays and find that the predictions of these models are vastly different from those of the SM. These features are common to NMSSM-like models with an arbitrary number of sterile superfields and the MSSM with broken R-parity.

## 1.Introduction

One of the most fundamental problems of particle physics is understanding the origin of CP-violation. In the Standard Model all CP-violating effects are described via Cabibbo-Kobayashi-Maskawa mechanism. In this approach, CP-violation originates from the quark mixing. However, in more general models, the origin of CP-violation can be quite different. For example, in multi Higgs Doublet Models, CP-symmetry can be broken spontaneously, that is to say, all CP-violating effects come entirely from complex phases in the Higgs VEV's [1]. In supersymmetric models, this approach allows to significantly reduce the number of free parameters and to aviod excessive CP-violation inherent in low energy supersymmetry [2]. In this work we discuss the CP phenomenology of these models which is quite different from what one expects in the Standard Model and, thus, allows to distinguish between the two approaches.

To begin with, we consider the NMSSM which has been shown to be the simplest acceptable susy model allowing for spontaneous CP-violation [3]. The Higgs neutral components develop the following VEV's:

$<H_1>= v_1$ , $<H_2>= v_2 e^{i\rho}$,
$<N>= n e^{i\xi}$ .

These complex phases enter the left-right squark and gaugino-higgsino mixings, as well as the quark masses. By a universal phase redefinition of the right-handed quarks and squarks, the quark masses as well as the gauge interaction vertices can be made real. Thus the effect of the complex VEV's shows up only in the Higgs (higgsino) sector and the left-right squark mixing.[1] This can lead to observable CP-violating effects in the quark sector through loop effects involving super-particles. It has been shown [2] that with a favorable choice of the parameters the model can predict correct values of $\epsilon$ and $\epsilon'$. In this work, we analyze the implications of the model for B physics.

## 2. CP-violation in B decays

One of the peculiar predictions of the Standard Model is the existence of the unitarity triangle. Information about the angles of this triangle can, for example, be extracted from the following decays [4]:

$$B \to \psi K_S \quad \sim \sin 2\beta \qquad (1)$$
$$B \to \pi^+ \pi^- \quad \sim \sin 2\alpha$$
$$B_S \to \rho K_S \quad \sim \sin 2\gamma$$

If the $B_L$-$B_H$ lifetime difference and the tree-penguin interference effects can be neglected, the rate of the B decays into a final CP-eigenstate is described by a simple formula:

$$\Gamma(B\to f_i) \propto e^{-\Gamma t} (1 - \sin 2\alpha_i \times \sin \Delta m t) \qquad (2)$$

Here $\alpha_i$ are the angles of the unitarity triangle. In the SM this relation is not exact since the interference between the tree and penguin contributions can be significant. One has to invoke isospin and SU(3) relations among different processes to separate these contributions and to determine the angles of the unitarity triangle more precisely. However, in susy models with spontaneous CP-violation, relation (2) is much more precise. This happens due to the fact that the CP-violating contribution comes from the superpenguin diagram (Fig.1) which is strongly suppressed by heavy propagators and loop factors. As a result, the interference between the tree and superpenguin contributions is negligible. It is important to note that, in our model, the angles of the unitarity triangle do not have a process independent meaning. In fact, they are defined by Eq.(2).

[1] In general, the left-right squark mixing cannot be redefined to be real by a universal phase transformation due to the presence of A-terms.

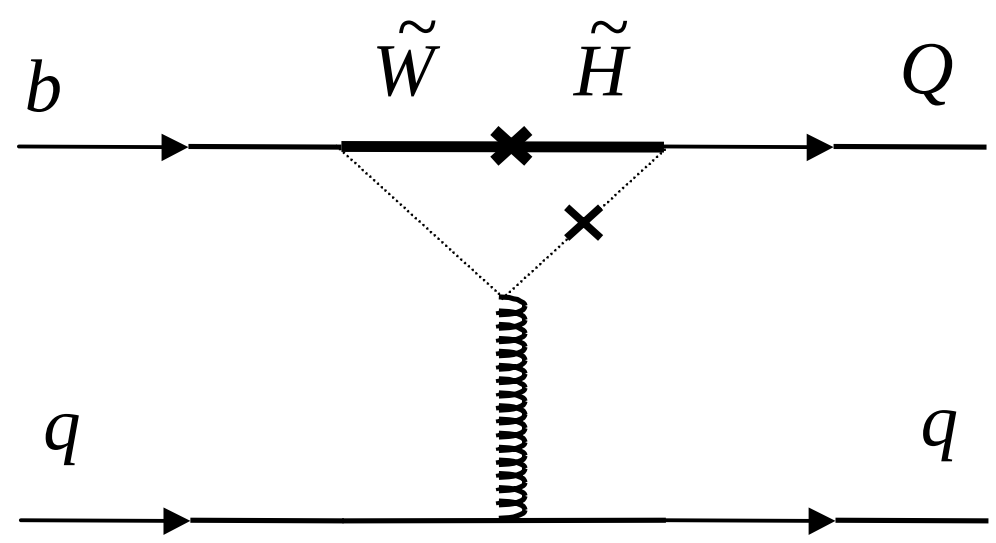

Fig.1.Dominant CP-violating contribution to B decays.

The angles of the unitarity triangle can be expressed in terms of the complex phases entering the mixing and decay diagrams:

$$\sin 2\alpha_i = \pm \sin(2\phi_D + \phi_M) \qquad (3)$$

As we have argued above, the decay phase $\phi_D$ is vanishingly small. The mixing phase $\phi_M$ can be considerably larger since both the SM and susy contribute at one loop level. In our model, there is a number of large CP-conserving contributions to $\Delta B=2$ operator such as the SM box, charged Higgs box, and the gluino and chargino superboxes. All of these contributions interfere constructively. On the other hand, all but the chargino CP-violating contributions (Fig.2) are suppressed by powers of $m_b / m_W$ [5]. In order to have a sizeable CP-violation in this model, the existence of relatively light susy particles is required.

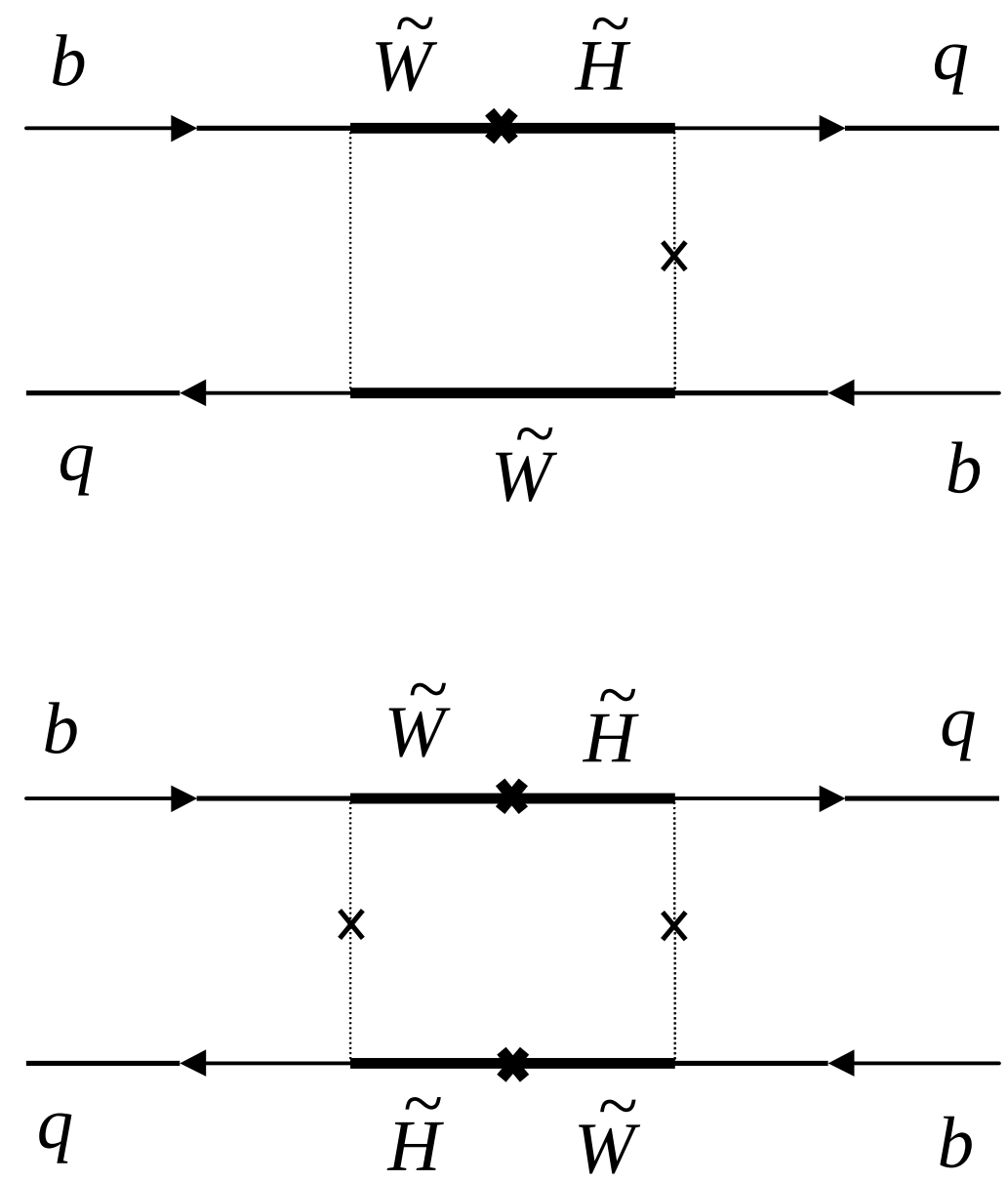

Fig.2. Dominant CP-violating contributions to the mixing (the cross on the scalar propagator denotes the L-R stop mixing).

Moreover, the NEDM bound constraints the phases quite severely. As a result, large CP-violation cannot be accommodated within such models.

From Eq.(3) one can determine typical values of the angles of the unitarity triangle. Since $\phi_M$ does not exceed 0.1 as required by the NEDM, the resulting triangle is either flat or very squashed and non-closing [6]. To be more quantitative, one can compare the prediction for the angle $\beta$ with its SM value. In order for the CKM approach to be consistent, $\sin 2\beta$ has to be greater than 0.4 (the CP asymmetry observed by CDF Collab. in B decays seems to agree with this prediction). It is interesting to see how large CP-violating phase is required in our model to reproduce this result and whether it is compatible with the NEDM bound. An illustrative plot showing parameter space regions allowed by sin2$\beta$≥0.4 and the NEDM is given in Fig.4. Apparently, the NEDM and sin2$\beta$≥0.4 cannot be accommodated at the same time. It can also be shown that the gap between the upper and lower bounds does not decrease considerably as one varies parameters of the model, i.e. tan$\beta$, chargino mass, GIM cancellation parameter, etc. [5].

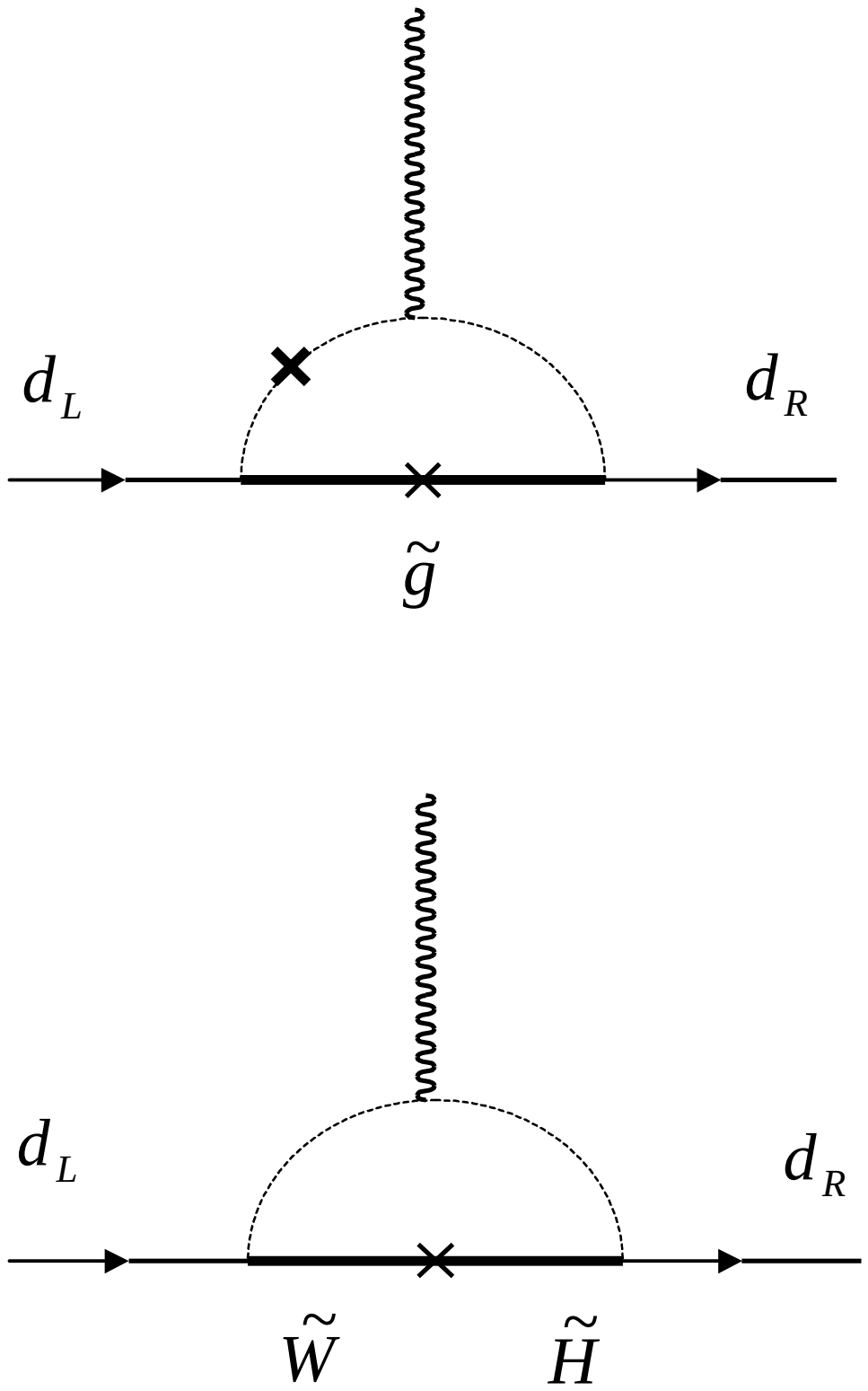

Fig.3. The NEDM diagrams allowing to constrain the phases entering the left-right squark and gaugino-higgsino mixings.

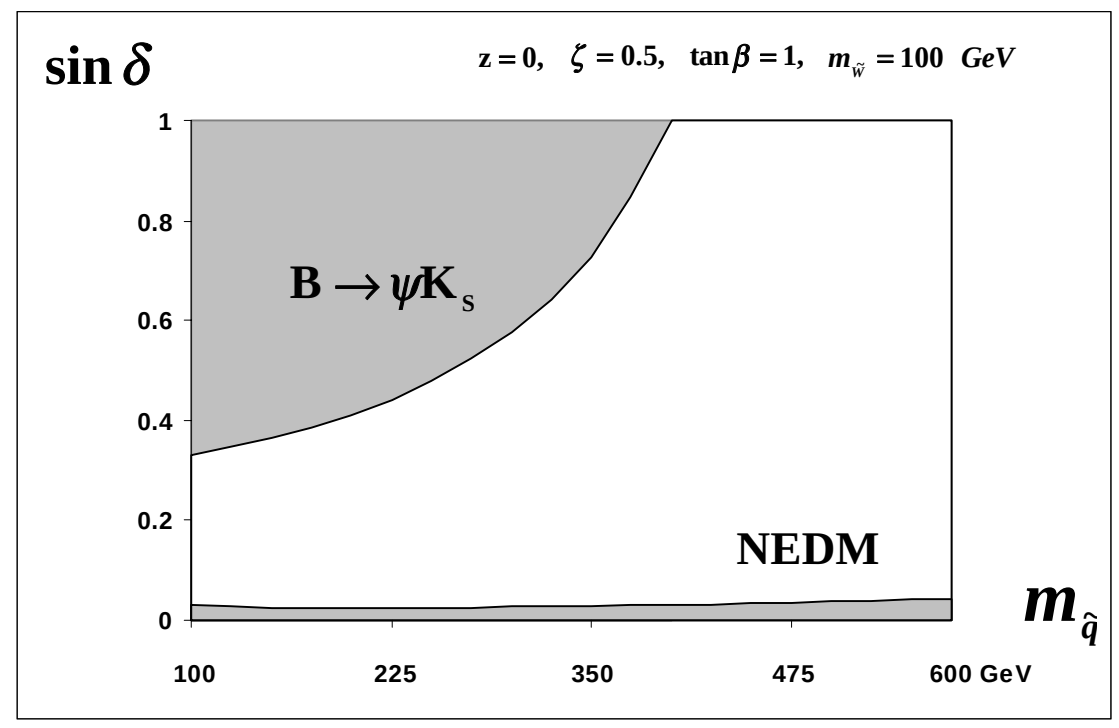

Fig.4. Regions of the parameter space allowed by sin2$\beta$≥0.4 and the NEDM.

In fact, the same considerations are valid for a more general situation – when one includes more than one sterile superfield. Even though there are many more phases in the scalar VEV's, these phases enter the NEDM and neutral meson mixing diagrams in the *same* combination.

Thus, if the SM prediction for sin2$\beta$ gets confirmed, a whole class of susy models with spontaneous CP-violation will be ruled out.

Another "next-to-minimal" model admitting spontaneous CPV is the MSSM with broken R-parity. In general, the sneutrinos can develop complex VEV's and, in this case, one effectively deals with a 5 Higgs Doublet Model. Besides that, the FCNC are generated at the tree level. The diagrams relevant to our considerations are given in Fig.5. They are required not to exceed their SM counterparts. Since we are interested in CP-violation in the down type quark sector, we may neglect the effect of complex sneutrino VEV's due to the strict constraints on the absolute values of the latter (~250 keV) [7] and $\lambda'$ couplings.[2] Then, in order to produce CP-violation, one would have to make left-right sparticle mixing insertions on the propagators. Due to the chiral structure of the R-breaking interactions, at least two such insertions are required on each propagator. As a result, CP-violating diagrams get suppressed by the fermion mass squared and can be neglected as compared to their CP-conserving counterparts. This means that large CP-violation in the B sector ($\sin 2\beta \geq 0.4$) cannot be produced and the model suffers from the same shortcoming as the ones described previously.

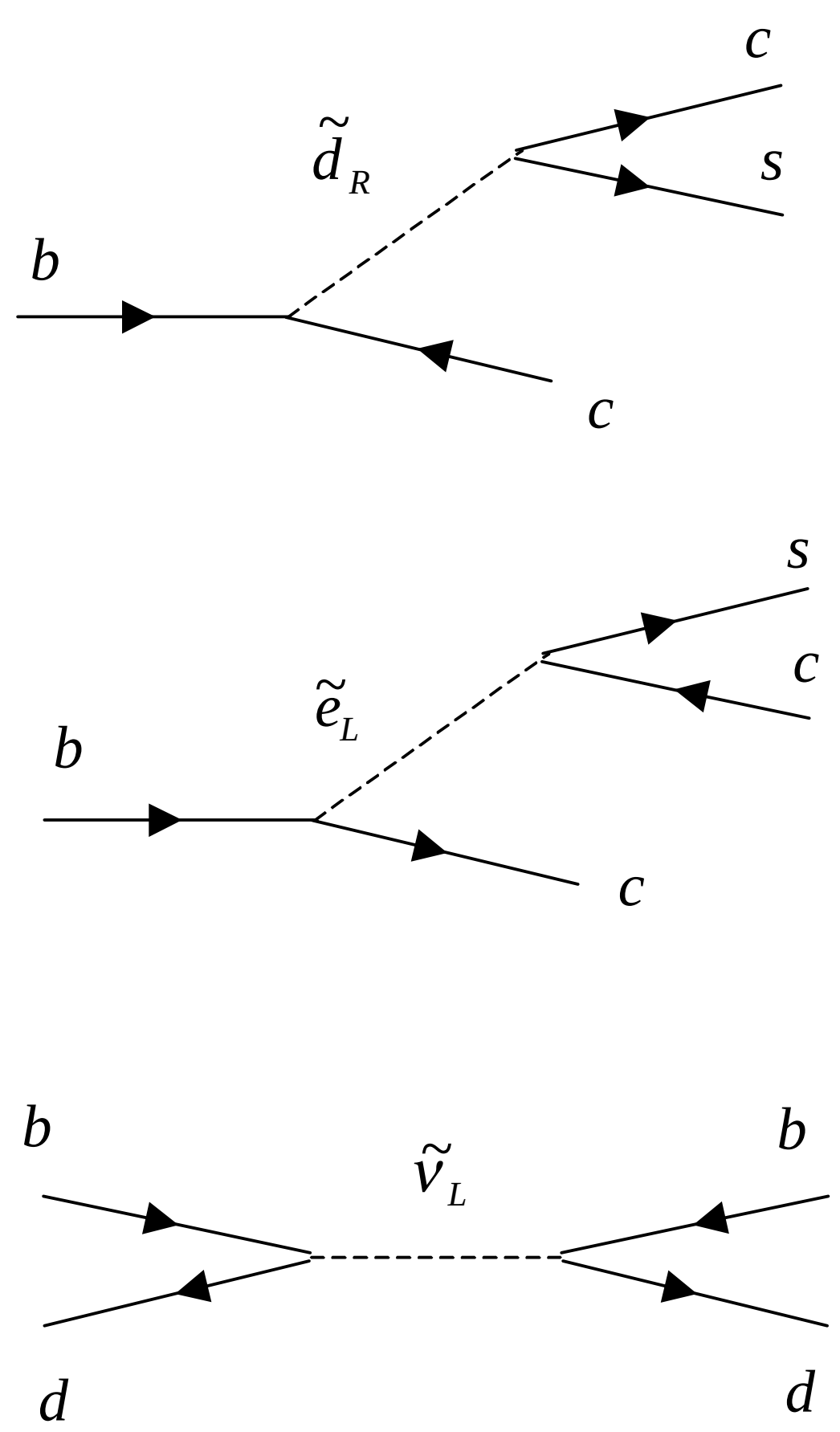

Fig.5. R-breaking contributions to $B \to \psi K_S$ decay and B-$\bar{B}$ mixing.

To summarize, the SM and susy models with spontaneous CP-violation lead to vastly different phenomenology in the B sector. The latter will be ruled out[3] if the SM prediction, $\sin 2\beta \geq 0.4$, is experimentally confirmed.

The author is grateful to the Fermilab Theory Group where part of this work was done.

[2] Via the $\lambda'$LQD interactions, the sneutrino VEV's may produce non-universal phases in the down type quark mass matrix. The effect, however, is small as explained above.

[3] We do not consider the possibility of accidental cancellations among contributions to the NEDM. The conclusions would not hold if such cancellations occurred.